\documentclass[preprint,12pt]{elsarticle}
\usepackage{amssymb}
\usepackage{multirow}
\usepackage{graphicx}
\usepackage{subfigure}
\usepackage{epsfig}
\usepackage[figuresright]{rotating}
\journal{Chemical Physics Letters}

\begin{document}

\begin{frontmatter}
\title{Lattice dynamics and electronic structure of energetic solids LiN$_3$ and NaN$_3$: A first principles study}
\author{K. Ramesh Babu and G. Vaitheeswaran}
\address{Advanced Centre of Research in High Energy Materials (ACRHEM),
University of Hyderabad, Prof. C. R. Rao Road, Gachibowli, Andhra Pradesh, Hyderabad- 500 046, India\\
\vspace{0.6in}
*Corresponding author E-mail address: gvsp@uohyd.ernet.in\\
\hspace {1.2in} Tel No.: +91-40-23138709\\
\hspace {1.2in} Fax No.: +91-40-23010227}
\end{frontmatter}
\newpage
\section*{Abstract} We report density functional theory calculations on the crystal structure, elastic, lattice dynamics and electronic properties of iso-structural layered monoclinic alkali azides, LiN$_3$ and NaN$_3$. The effect of van der Waals interactions on the  ground- state structural properties is studied by using various dispersion corrected density functionals. Based on the equilibrium crystal structure, the elastic constants, phonon dispersion and phonon density of states of the compounds are calculated. The accurate energy band gaps are obtained by using the recently developed Tran Blaha-modified Becke Johnson (TB-mBJ) functional and found that both the azides are direct band gap insulators. \\
Keywords:
van der Waals interactions; elastic constants; phonon dispersion; electronic structure
\newpage
\section{Introduction}
Inorganic metal azides are an important class of energetic materials that find numerous applications as explosives and gas generators \cite{Evans, Fair, Bowden}. Among the inorganic metal azides, alkali metal azides LiN$_3$ \cite{MZhang} and NaN$_3$ \cite{Yin, Hou} received much interest in recent years because of their ability to form polymeric nitrogen, a green high energy density material, at extreme conditions. In particular, by using NaN$_3$ as starting material, Eremets et al. \cite {Eremets} have synthesized the polymeric networks of nitrogen by combined high pressure X-ray powder diffraction and Raman spectroscopy techniques. Their study concluded that double bonded N atoms of the azide ion become single bonded N atoms at a pressure of around 120 GPa. The recent high pressure experiments on LiN$_3$ confirms that the material has the ability to form single bonded networks of nitrogen beyond the compression of 62 GPa \cite{Medvedev}. In addition, the typical crystal structures of LiN$_3$ and NaN$_3$ which contain molecular azide anion units, serve as model systems to study the fast chemical reactions that occur in complex molecular solids.
\paragraph*{} Lithium azide and the low temperature phase of sodium azide are iso-structural and crystallize in monoclinic structure with \textit{C2/m} space group \cite{Pringle}. The unit cell contains two formula units with metal cations occupies the site with fractional co-ordinates (0, 0, 0), central nitrogen atoms of the azide occupies the (0 0.5 0.5) site whereas the end nitrogen atom is situated at (0.1048, 0.5, 0.7397) for LiN$_3$ and at (0.1016, 0.5, 0.7258) for NaN$_3$ respectively. There are few theoretical studies available on the electronic structure \cite{Seel, Ju, Gord, Ram} and optical properties \cite{Zhu, Zhu-1, Ramesh} of the metal azides. The elastic constants of LiN$_3$ were reported by using the standard density functional calculations \cite{Ramesh, Perger}. But, both the metal azides have layered crystal structure environment as shown in Fig. 1, with the presence of van der Waals (vdW) interactions between the layers. Therefore the inclusion of vdW interactions in the calculations is must for well description of crystal structure and related properties of the systems. Although density functional theory (DFT) is the robust theoretical approach to predict the material properties but the conventional DFT techniques based on local density approximation (LDA) and generalized gradient approximation (GGA) will tend to give large errors when compared to experiments due to the lack of description of vdW interactions \cite{Jones, Bucko, Bheem, KRB, KRamesh, Bheem2, Appa, Appa1, Appa2}. To the best of our knowledge, the dispersion corrected density functional theory based calculations on these metal azides to explore the elastic and phonon properties were not performed yet. In this work, we have carried out the study of effect of vdW interactions on the crystal structure and thereby the elastic properties, Born-effective charges, and phonon dispersion of the metal azides in order to understand the mechanical and dynamical stability.
\paragraph*{} It is well known that the metal azides undergo decomposition in to metal atom and nitrogen when irradiated with the light of energy greater than that of the optical band gap value \cite{Evans}. Therefore the knowledge of exact band gap value would be of useful to understand the decomposition phenomena of the azides. To the best of our knowledge there are no experimental band gap available for these compounds and also on the other hand the usual density functional calculations result in underestimation of band gap by 30-40 $\%$ \cite{Payne}. To overcome the band gap problem, recently developed TB-mBJ functional has been widely used in the literature \cite{Cam} as it takes less computational time when compared to the very expensive GW calculations \cite{Bheem1}. Hence, in this study we have also carried out the energy band structure calculations using the recently developed functionals to get the exact band gap value of the azides. To know the optical response of the azides we have calculated the absorptive part of the complex dielectric function. The remainder of the paper is organized as follows: A brief description of our computational details is presented in section 2. The results and discussion are presented in section 3 followed by summary of our conclusions in  section 4.
\section{Computational details}
The calculations were performed using plane wave pseudo potential method based on density functional theory \cite{Segall}. The interactions between the ions and electrons are described by using Vanderbilt ultrasoft pseudo potentials \cite{Vanderbilt}. For all the calculations we have included the 2s$^1$ electrons for Lithium, 3s$^1$ electrons for Sodium and the 2s$^2$, 2p$^3$ states of nitrogen. Both local density approximation (LDA) of Ceperley and Alder \cite{Ceperley} parameterized by Perdew and Zunger (CA-PZ) \cite{PPerdew}  and also the generalized gradient approximation (GGA) with the Perdew-Burke-Ernzerhof (PBE) \cite{Perdew} parameterizations were  used for the exchange-correlation potentials. The calculations were performed using an energy cut-off of 520 eV for the plane wave basis set. Integrations in the Brillouin zone were performed according with a 5x8x5 Monkhorst-Pack grid scheme \cite{Monkhorst} k-point mesh. The changes in the total energies with the number of k-points and the cut-off energy were tested to ensure the convergence within one meV per atom.
\paragraph*{} To treat vdW interactions efficiently, we have used the vdW correction to the exchange - correlation functional of standard density functional theory at semi empirical level. According to semi-empirical dispersion correction approach, the total energy of the system can be expressed as
\begin{equation}
E_{total} = E_{DFT} + E_{Disp}
\end{equation}
where
\begin{equation}
E_{Disp} = s_i\Sigma_{i=1}^N\Sigma_{j>i}^Nf(S_RR^{0}_{ij}, R_{ij})C_{6, ij}R_{ij}^{-6}
\end{equation}
here C$_{6, ij}$ is called dispersion coefficient between any atom pair $i$ and $j$ which solely depends upon the material and R$_{ij}$ is the distance between the atoms $i$ and $j$. In the present study we have used the recently developed schemes by Ortmann, Bechstedt, and Schmidt \cite{OBS} in LDA and Grimme \cite{Grimme}, Tkatchenko - Scheffler \cite{TS} approaches in GGA. These semiempirical approaches provide the best compromise between the cost of first principles evaluation of the dispersion terms and the need to improve non-bonding interactions in the standard DFT description. For the computation of electronic properties, we have used the linearized augmented plane wave (LAPW) method as implemented in WIEN2k package \cite{DJ, PBlaha}. We have used Engel-Vosko functional \cite{EV} and recently developed Tran Blaha-modified Becke Johnson potential \cite{TB} within GGA to get the accurate band gaps of the compounds in addition to the usual CA-PZ and PBE functionals.
\section{Result and discussion}
\subsection{Structural properties}
The crystal structure of LiN$_3$ and NaN$_3$, as shown in Fig. 1, consists of the axes of the azide ions parallel to each other and almost perpendicular to the layer in which they are stacked. In our previous work on lithium azide, LiN$_3$ \cite{Ram, Ramesh} we used three different functionals (LDA (CA-PZ), GGA (PBE) and GGA (PBE+G06)) to predict the theoretical equilibrium crystal structures. For better description of the azide systems, in this present study, we have used recently developed dispersion corrected functionals such as LDA (OBS), GGA (PBE+TS) in addition to the LDA (CA-PZ), GGA (PBE) and GGA (PBE+G06) functionals and studied their virtue in describing the crystal structure of LiN$_3$ and NaN$_3$. As a first step, we have optimized the crystal structures with the conventional LDA (CA-PZ) and GGA (PBE) functionals. The calculated structural parameters using both the functionals are presented in Table 1. The crystal volume computed is under estimated by -11.4 $\%$ and -9.5 $\%$ respectively for LiN$_3$ and NaN$_3$ in the CA-PZ based computation. Whereas the PBE volume is overestimated by 5.7 $\%$ for LiN$_3$ and 7.9 $\%$ for NaN$_3$. The errors in computed equilibrium volume can be attributed to the fact that both functionals are inadequate to treat the weak dispersive forces present between the azide ions. To treat the vdW interactions efficiently, we have performed the crystal structure optimization by using the dispersion corrected functional such as CA-PZ+OBS in LDA and PBE+TS and PBE+G06 in GGA. The results obtained are presented in Table 1 along with the LDA and GGA results. Among the three dispersion corrected functionals considered in this work, the best qualitative agreement between the experiment and theory for the lattice parameters is achieved by PBE+G06 functional. The equilibrium volume of LiN$_3$ (NaN$_3$) is under estimated by -19.1 $\%$ (-15.6 $\%$) within CA-PZ+OBS functional, -4.3 $\%$ (-0.8 $\%$) within PBE+TS functional and by -1.8 $\%$ (-0.08 $\%$) using PBE+G06 functional. Therefore, we use the PBE+G06 functional for further calculations of elastic constants and phonon properties which are strongly related to the optimized crystal structures.
\subsection{Bulk modulus and its pressure derivative}
The bulk modulus, B$_0$, and its pressure derivative B$_0$$^\prime$ can be deduced from the pressure and volume data of the crystal lattice. In the present study, the calculated pressure-volume data of LiN$_3$ and NaN$_3$ using the PBE+G06 functional are fitted to the Murnaghan's equation of state. The bulk modulus and its pressure derivative for LiN$_3$, are obtained to be 36.3 GPa and 4.108, respectively whereas for NaN$_3$ the calculated values are 16.3 GPa and 5.692. When compared with the experimental value of B$_0$ = 19.1$\pm$ 1.4 GPa and B$_0$$^\prime$ = 7.3$\pm$ 0.5 \cite{Medvedev}, the calculated value of B$_0$ for LiN$_3$ is higher which can be attributed to the fact that the equilibrium volume is lower by -1.8 $\%$ compared to experiment. The present theoretical values of B$_0$ and B$_0$$^\prime$ for NaN$_3$ are in good comparison with the experimental values of 17.56 $\pm$ 0.7GPa and 5.66 $\pm$ 0.2 \cite{Hou}.
\subsection{ Elastic constants}
According to Hooke's law, for a given small strain applied to the lattice the stress components can be calculated using,
\begin{equation}
\sigma_{ij} = C_{ijkl} \epsilon_{kl}
\end{equation}
where $\sigma$$_{ij}$ is the stress tensor, $\epsilon$$_{kl}$ is the strain tensor and C$_{ijkl}$ is the elastic stiffness tensor. By following the Voigt's notation\cite{Nye}, this equation can be reduced to
\begin{equation}
\sigma_{i} = C_{ij} \epsilon_{j} \hspace{0.5cm} with \hspace{0.5cm} i=1-6, j= 1-6
\end{equation}
where $xx, yy, zz, yz, xz, xy$ replaced by 1, 2, 3, 4, 5, 6 respectively and therefore C$_{ij}$ form a 6x6 matrix.
 Due to the structural symmetry, the maximum number of independent parameters can be reduced to thirteen for the monoclinic phase. We have calculated the thirteen independent elastic constants of monoclinic LiN$_3$ and NaN$_3$ and are presented in Table 2. Clearly, the calculated elastic constants follow the Born-Huang mechanical stability criterion for a monoclinic system \cite{Born} indicating that both the compounds are mechanically stable systems. In general, for a monoclinic lattice, C$_{11}$, C$_{22}$ and C$_{33}$ denote the elastic properties along the a, b and c-axes of the lattice respectively. It can be seen that there is a considerable elastic anisotropy among the three principal directions due to the fact that C$_{11}$ $\neq$ C$_{22}$ $\neq$ C$_{33}$ for both the azides. Moreover, we find that C$_{22}$ value is lower than that of the other two elastic constants C$_{11}$ and C$_{33}$ for both the metal azides. This implies that both the lattices are less stiffer along the b-axis and can easily undergo deformation under the applied stress when compared to the a-and c-axes. This result is in good accord with the recent experimental finding that the strongest deformation occurs along b-axis of LiN$_3$ under the application of pressure \cite{Medvedev}.
\subsection{Phonon dispersion and phonon density of states}
The lattice dynamical properties were calculated within the framework of density functional perturbation theory (DFPT) \cite{Gonze, Refson}. All the calculations were carried out by using norm-conserving pseudo potentials available in CASTEP code with an energy cut-off of 770 eV. For all the calculations we have used  5x8x5 grid of k-points for the Brillouin zone integration. The primitive unit cell of the metal azides contains four atoms giving rise to a total of 12 phonon branches. The \textit{C2/m} space group, which describes the monoclinic symmetry, has four irreducible representations namely, A$_u$, B$_u$, A$_g$ and B$_g$. A group-theoretical analysis gives the following decomposition of vibrational representation into its irreducible components at the $\Gamma$-point: $\Gamma$-tot = 2A$_g$ + 1B$_g$ + 3A$_u$ + 6B$_u$. Out of these modes, three (A$_u$ + 2B$_u$) are acoustic modes and 9 are optical modes given by $\Gamma$$_{opt}$ = 2A$_g$ + 1B$_g$ + 2A$_u$ + 4B$_u$. In these, A$_u$ and B$_u$ are infrared active modes and the remaining are Raman active modes.
\paragraph{} The calculated phonon dispersion spectrum and the phonon density of states of LiN$_3$ and NaN$_3$ are presented in Fig. 2 and 3, respectively. The phonon dispersion of the metal azides is plotted along the high symmetry points of the Brillouin zone namely, (0 0 0) $\Gamma$ $\rightarrow$ (-0.5 -0.5 -0.5) M $\rightarrow$ (-0.5 0 0) A $\rightarrow$ (0 0 0) $\Gamma$ $\rightarrow$ (0 -0.5 0.5) Z $\rightarrow$ (-0.5 0 0.5) L.
From the phonon dispersion spectra, one can notice that at low energies of the gamma point, there are three acoustic modes out of which one is longitudinal (LA) mode and the remaining two are degenerate transverse (TA) modes. All these modes have linear dispersion at the lower energy. The infrared-active optical modes, (A$_u$ and B$_u$) at $\Gamma$-point split into longitudinal optic (LO) and transverse optic (TO) components with different frequencies, which results in discontinuities of the phonon branches in the Brillouin zone centre of the phonon dispersion curves of LiN$_3$ and NaN$_3$. The origin of these effects is the long-range macroscopic electric field accompanying atomic displacements \cite{Bar}. The value of LO/TO splitting is determined by the Born effective charge tensors Z$^*$$_{k, \alpha \beta}$ and high-frequency dielectric constant tensor $\epsilon^\infty$ via a nonanalytic contribution to the force constants. The calculated values of these quantities are tabulated in Table 3. The Born effective charge tensor, Z$^*$$_{k, \alpha \beta}$, is a fundamental quantity for the study of lattice dynamics and are defined as the force in the direction $\alpha$ on the atom k due to an homogeneous electric field along the direction $\beta$ or equivalently as the induced polarization of the solid along the direction $\alpha$ by a unit displacement in the direction  $\beta$ of the atomic sub-lattice. These can be further used to describe the degree of covalency or ionicity in a crystal. The calculated Born effective charges of the metal atoms Z$^*$ (Li) and Z$^*$ (Na) are close to the nominal charge of Li and Na atoms (+1), indicating that both the atoms involve in an ionic type interaction. The deviation of the calculated Born effective charges of mid N and end N from their formal charges -1 and -2 implies that the charge sharing is dominating over N-N bond. Effective charge tensors of mid N and end N atoms have significant off-diagonal components, as well as display anisotropy of their diagonal terms. The largest diagonal term inequality is found for the end N atoms. The mean values of the diagonal components of end N effective charge tensors are 0.94 for LiN$_3$ and 0.96 for NaN$_3$, respectively. This implies that NaN$_3$ is characterized by higher dynamic ionicity compared to LiN$_3$.
\paragraph{} In Table 4 we have summarized the computed frequencies of the TO and LO optic modes of Au and Bu symmetry. The largest splitting (~22cm$^{-1}$) is found in the Au mode. It can be noticed from the calculated phonon dispersion spectra of the azides that LO/TO splitting is not much pronounced because of the weak interaction between the metal cation and the azide anion. The phonon branch at higher energies i.e., at around 2000 cm$^{-1}$ is due to the asymmetric stretching of the azide ion and it is the infrared active Bu mode. The Raman active stretching mode (A$_g$) of the azide ion is located at around 1300 cm$^{-1}$. The doubly degenerate infrared active A$_u$ mode, which is entirely due to the symmetric bending of the azide ion is situated at 620 cm$^{-1}$. These calculated frequencies at the gamma point are in good comparison with that of measured frequencies \cite{Evans, Massa}. By looking at the polarization vector of each mode we noticed that all the higher energy phonon branches are due to the motion of the azide anion sublattice and therefore these modes can be classified as internal modes. The phonon branches that are situated at lower energies up to about 350 cm$^{-1}$ for LiN$_3$ and 250 cm$^{-1}$ for NaN$_3$ are due to the vibrations from the total lattice and can be called as lattice modes or external modes. There are five external mode phonon branches out of which four modes are infrared active and the only Raman active mode situated at 120 cm$^{-1}$ due to the asymmetric bending of azide ion sub-lattice. The present calculations with PBE+G06 functional could reproduce the frequencies of the lattice modes which are in good comparison with the experiments \cite{Evans, Massa}.
\section{Electronic structure and optical properties}
\paragraph*{} Both LiN$_3$ and NaN$_3$ are relatively stable at normal conditions when compared to other metal azides. However, when irradiated with suitable wave length of light, both the metal azides show instability and undergo decomposition into metal and nitrogen atoms respectively \cite{Evans, Fair}. Thus, it is very important to know about the exact band gap energies of the azides to understand the optical response of the materials. Previously, the band gap values were reported within GGA to be 3.7 eV \cite{MZhang}, 3.46 eV \cite{Gord}, 3.32 eV \cite{Ram, Ramesh}, 4.684 eV \cite{Zhu-1} for LiN$_3$ and 5.034 eV for NaN$_3$ \cite{Zhu-1}. However these band gap values might be underestimated by 30-40 $\%$ \cite{Payne} from experiment due to the inherent band gap problem with usual DFT functionals. To overcome the band gap problem, we have calculated the band gap energies of LiN$_3$ and NaN$_3$ by using TB-mBJ functional in addition to the other exchange-correlation functionals. It is found that both the metal azides are direct band gap insulators (Z-Z) with a gap of 3.801 eV (LDA), 3.981 eV (GGA), 4.392 eV (EV), 4.981 eV (TB-mBJ) for LiN$_3$ and 4.171 eV (LDA), 4.023 eV (GGA), 4.272 eV (EV), 5.380 eV (TB-mBJ) for NaN$_3$. The computed band structures within GGA with TB-mBJ band gap energies are shown in Fig. 4 (a) and 4 (b) respectively. There are no experimental band gap values to compare the present band gap values. However, it is well known that the TB-mBJ functional provides the band gap values that are in close agreement with the experiments \cite{Cam, TB}, we expect our present values will be useful for further studies on these compounds. The imaginary part of the complex di-electric function of the metal azides have been calculated by using the TB-mBJ functional and are shown in Fig. 4 (a) and 4 (b) respectively. The computed absorption part of the complex di-electric function of both the metal azides show similar features up to 20 eV. For both the metal azides, the optical absorption starts at the band gap values. As the optical edge starts at 4.981 eV in LiN$_3$ and at 5.380 eV in NaN$_3$ we conclude that lithium azide will undergo decomposition by the irradiation of UV light with wave length of 248.9 nm and sodium azide with that of 230.5 nm.
\section{Conclusion}
In conclusion we have studied the effect of van der Waals interactions on the structural properties of the metal azides LiN$_3$ and NaN$_3$. We have examined the virtues of various dispersion corrected functionals. We found that PBE+G06 functional gives accurate structural results when compared to CA-PZ+OBS and PBE+TS dispersion corrected functionals. The optimized crystal structure with the PBE+G06 functional was used for the calculation of elastic constants and the phonon properties such as phonon dispersion and phonon density of states. From the calculated elastic constants we conclude that both the metal azides are mechanically stable systems. The phonon dispersion and the phonon density of states of LiN$_3$ and NaN$_3$ were calculated and found that both the metal azides are dynamically stable systems as there is no imaginary phonons. The phonon branches were analyzed and the calculated phonon frequencies at the gamma point are in good agreement with experiments. The energy band gaps were calculated using TB-mBJ functional and also optical response of the materials were analyzed through the absorptive part of the complex dielectric function.
\section{ACKNOWLEDGMENTS}
K R B would like to acknowledge DRDO for funding through ACRHEM and also CMSD, University of Hyderabad for computational facilities. G V and K R
B thank Prof. M. C. Valsa Kumar, School of Engineering Sciences and Technology, University of Hyderabad for critical reading of the manuscript and fruitful discussions.

\newpage
\textbf{Figure Legends:} \vspace{1cm}  \\
Figure 1: (Colour Online) Crystal structure of MN$_3$ (M= Li or Na). In the figure, violet ball indicates metal atom and blue ball indicates nitrogen atom, respectively. \vspace{1cm} \\

Figure 2: (Colour online) Phonon dispersion and phonon density of states of LiN$_3$ (a) and NaN$_3$ (b) calculated at theoretical equilibrium volume within PBE+G06 functional \vspace{1cm}\\


Figure 3: (Colour online) Electronic band structure of  LiN$_3$ (a) and NaN$_3$ (b) calculated at experimental volume within GGA by using the TB-mBJ functional. \vspace{1cm}\\

Figure 4: (Colour online) Imaginary part of the directionally averaged macroscopic dielectric function of  LiN$_3$ (a) and  NaN$_3$ (b) calculated at experimental volume with TB-mBJ functional.

\clearpage
\begin{figure}
\centering
\includegraphics[width=100mm, height=100mm]{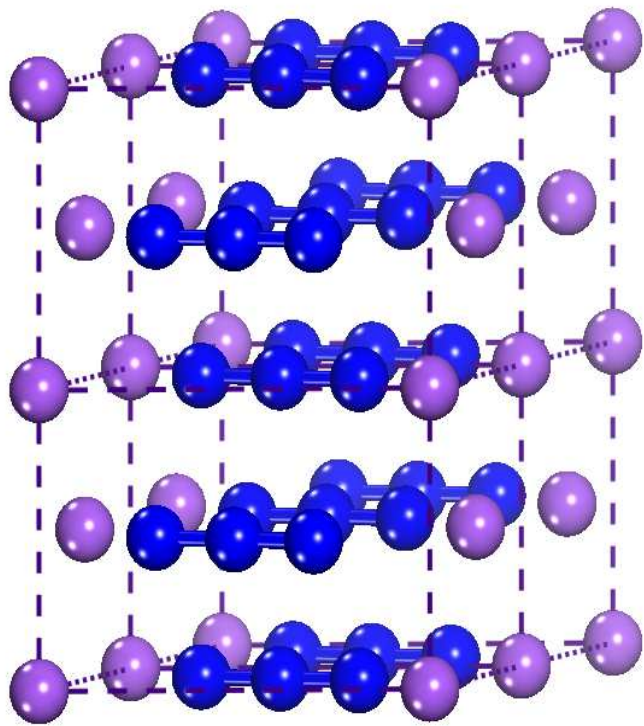}\\
\caption{(Colour Online) Crystal structure of MN$_3$ (M= Li or Na). In figure, violet ball indicates, metal atom and blue ball indicates nitrogen atom, respectively.}
\end{figure}

\begin{figure}
\begin{center}
\subfigure[]{\includegraphics[width=140mm,height=70mm]{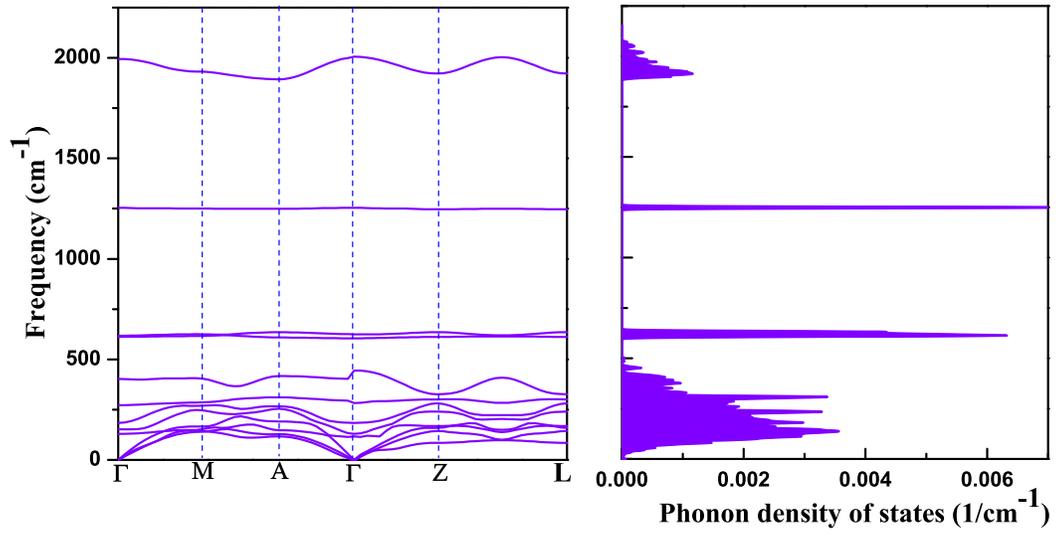}}
\subfigure[]{\includegraphics[width=140mm,height=70mm]{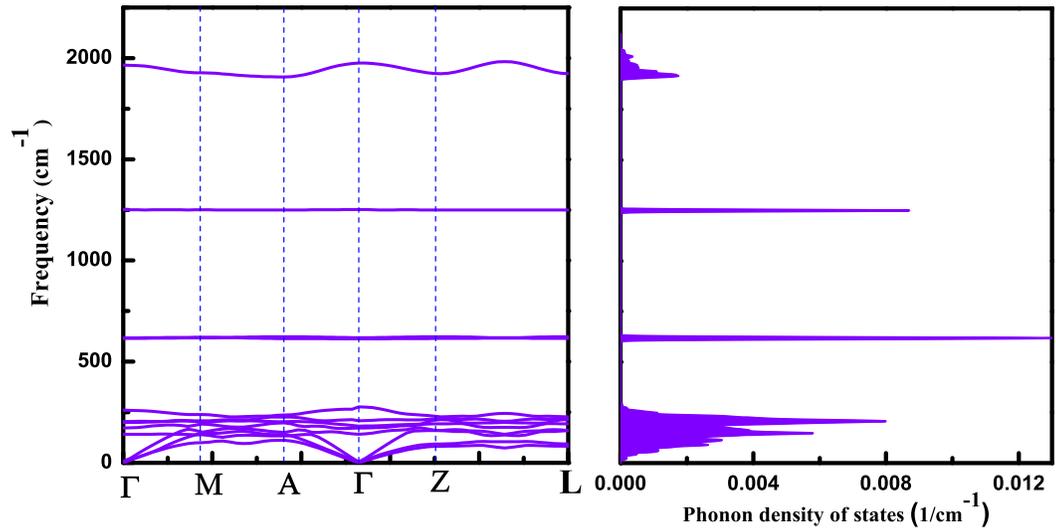}}\\
\caption{(Colour online) Phonon dispersion and phonon density of states of LiN$_3$ (a) and NaN$_3$ (b) calculated at theoretical equilibrium volume within PBE+G06 functional}
\end{center}
\end{figure}

%

\begin{figure}
\begin{center}
\subfigure[]{\includegraphics[width=65mm,height=80mm]{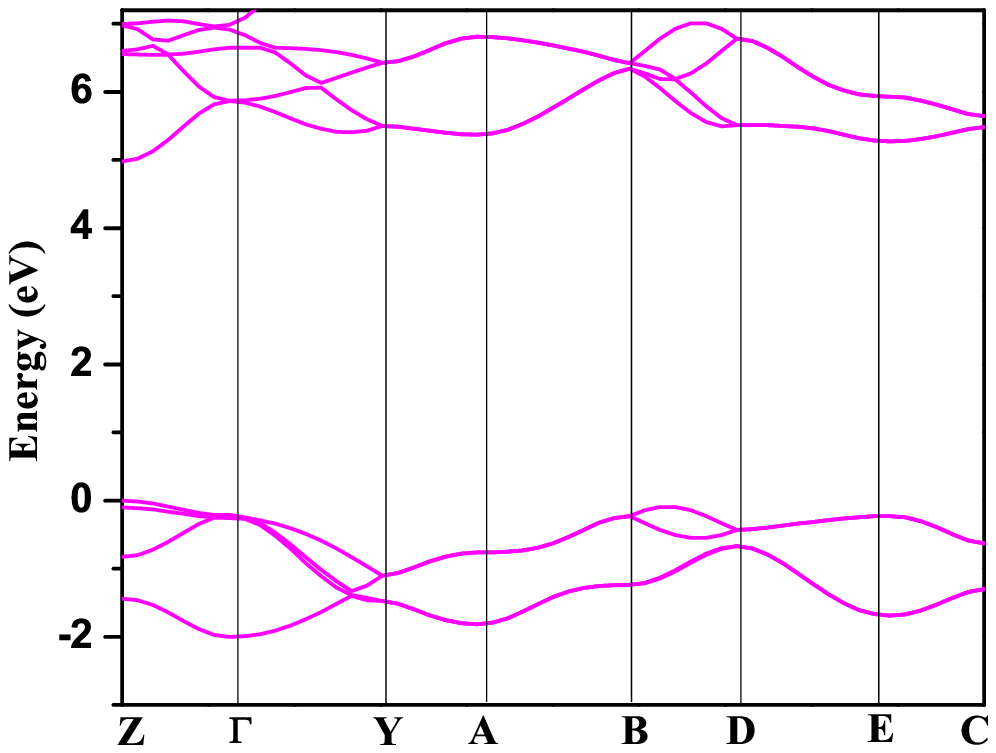}}
\subfigure[]{\includegraphics[width=65mm,height=80mm]{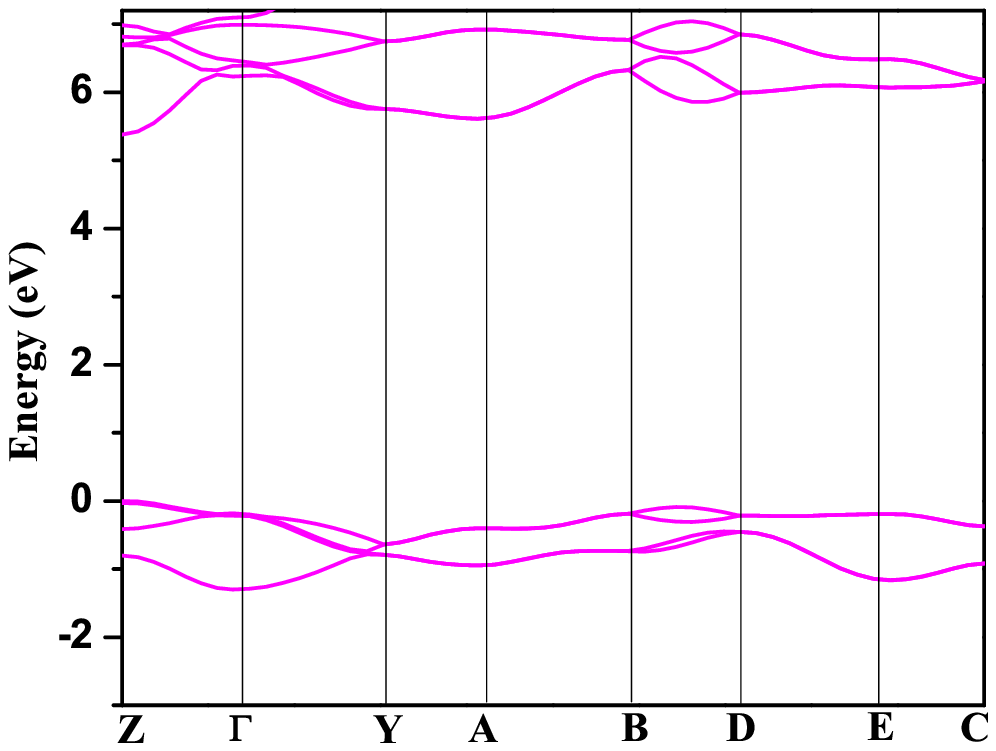}}\\

\caption{(Colour online) Electronic band structure of  LiN$_3$ (a) and NaN$_3$ (b) calculated at experimental volume within GGA by using the TB-mBJ functional.}
\end{center}
\end{figure}

\clearpage
\newpage
\begin{figure}
\begin{center}
\subfigure[]{\includegraphics[width=100mm,height=70mm]{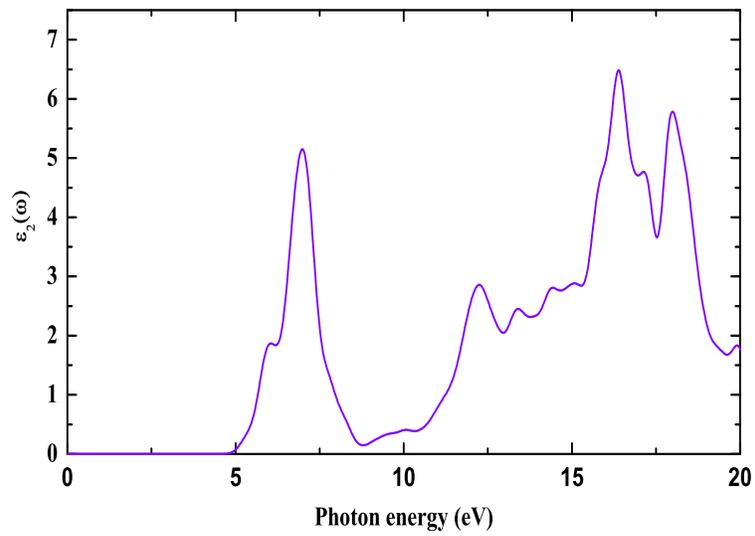}}
\subfigure[]{\includegraphics[width=100mm,height=70mm]{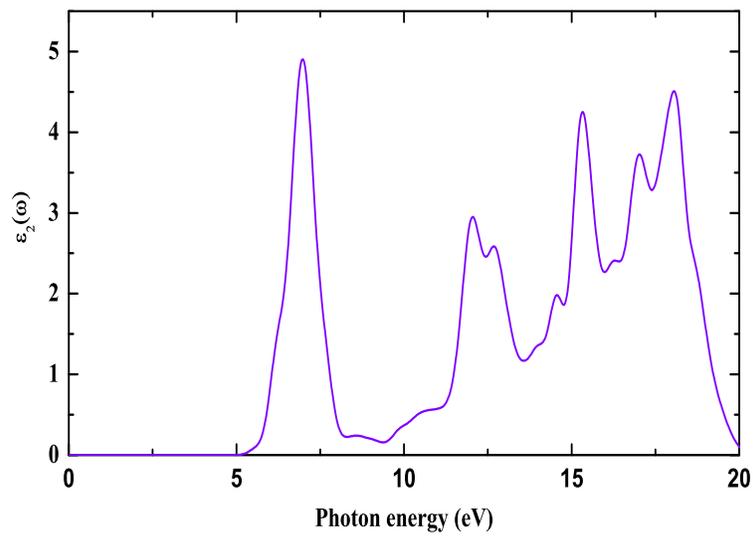}}\\
\caption{(Colour online) Imaginary part of the directionally averaged macroscopic di-electric function of  LiN$_3$ (a) and  NaN$_3$ (b) calculated at experimental volume with TB-mBJ functional.}
\end{center}
\end{figure}
\clearpage
\newpage
\begin{table}[ht]
\caption{ Ground state properties of monoclinic LiN$_3$ and NaN$_3$ calculated using various exchange-correlation functionals.}
\begin{tabular}{ccccccccc} \hline
LiN$_3$ & & &  & & & \\
XC functional & a (\AA) & b (\AA) & c (\AA) & $\beta$$^0$ & N & V (\AA$^3$) \\ \hline \hline
CA-PZ & 5.328&3.199 &4.727 &102.6 & (0.1244 0.5 0.7456)& 78.6\\
CA-PZ+OBS & 5.158& 3.074 & 4.615 & 100.9 & (0.1288 0.5 0.7459) & 71.8 \\
PBE & 5.761 & 3.376 & 5.094 & 108.6 & (0.1005 0.5 0.7490)& 93.9 \\
PBE+TS & 5.541 & 3.296 & 4.796 & 104.3 & (0.1174 0.5 0.7490) & 84.9 \\
PBE+G06 & 5.686 & 3.219 & 4.884 & 102.7 & (0.1094 0.5 0.7257) & 87.2\\
Expt & 5.627 & 3.319 &4.979 & 107.4 & (0.1048 0.5 0.7397) & 88.8 \\ \hline \hline
NaN$_3$ & & &  & & & \\
XC functional & a (\AA) & b (\AA) & c (\AA) & $\beta$$^0$ & N & V (\AA$^3$) \\ \hline \hline
CA-PZ & 5.868&3.568 &5.081 &102.5 & (0.1148 0.5 0.7277)& 103.8\\
CA-PZ+OBS & 5.670& 3.483 & 4.981 & 100.2 & (0.1219 0.5 0.7235) & 96.8 \\
PBE & 6.446 & 3.748 & 5.473 & 110.3 & (0.0930 0.5 0.7350)& 123.9 \\
PBE+TS & 6.243 & 3.616 & 5.269 & 106.8 & (0.1055 0.5 0.7329) & 113.8 \\
PBE+G06 & 6.217 & 3.647 & 5.229 & 104.7 & (0.1094 0.5 0.7257) & 114.7 \\
Expt & 6.211 & 3.658 & 5.323 & 108.4 & (0.1016 0.5 0.7258) & 114.8 \\ \hline \hline
\end{tabular}\\

\end{table}
\clearpage

\newpage
\begin{table}[ht]
\caption{Single crystal elastic constants (C$_{ij}$, in GPa) of monoclinic LiN$_3$ and NaN$_3$ calculated within PBE+G06 functional}
\begin{tabular}{cccccccccccccc} \hline \hline
 Compd& C$_{11}$&C$_{22}$&C$_{33}$&C$_{44}$&C$_{55}$&C$_{66}$&C$_{12}$&C$_{13}$&C$_{15}$&C$_{23}$&C$_{25}$&C$_{35}$&C$_{46}$\\ \hline
LiN$_3$&92.2 & 77.3& 116.6 & 13.3 & 36.3 & 22.3 & 21.4 & 37.9 & 19.9 & 30.2 & 2.9 & 41.4 & -6.1\\
NaN$_3$&64.6 & 31.2 & 62.1 & 7.2 & 21.9 & 8.6 & 10.8 & 27.8 & 21.3 & 16.5 & 1.9 & 18.6 & -2.9 \\
\hline
\end{tabular}\\
\end{table}
\clearpage
%
%
\clearpage
\newpage
\begin{table}
\caption{The Born-effective charges and high frequency dielectric constants of monoclinic LiN$_3 $ and NaN$_3$ calculated within GGA-PBE+G06 functional.}
\label{tab.1}
\begin{center}
\begin{tabular}{ccc}\hline \hline
 & LiN$_3$ & NaN$_3$ \\ \hline
Z$^*_{xx}$ &  Li: 1.18  & Na: 1.14  \\
           & mid N: -0.98  & mid N: -1.02  \\
           & end N: 0.78 & end N: 0.89 \\
           & & \\
Z$^*_{yy}$ &  Li: 1.12 & Na: 1.05 \\
           & mid N: -0.57  & mid N: -0.51\\
           & end N: -0.02 & end N: -0.04 \\
           & & \\
Z$^*_{zz}$ &  Li: 1.78 & Na: 1.47 \\
           & mid N: -1.93  & mid N: -1.76\\
           & end N: 2.08& end N: 2.04 \\
           & & \\
Z$^*_{xz}$ &  Li: 0.15 & Na: 0.14 \\
           & mid N: -1.93 & mid N: -0.79 \\
           & end N: 1.30 & end N: 1.44 \\
           & & \\
Z$^*_{zx}$ &  Li: 0.40 & Na: 0.29  \\
           & mid N: -0.76 & mid N:-0.79 \\
           & end N: 1.13 & end N: 1.30\\
$\epsilon_{xx}^\infty$ &2.45   & 2.42  \\
$\epsilon_{yy}^\infty$ &  2.08 & 1.95   \\
$\epsilon_{zz}^\infty$ & 3.75  &  2.99  \\   \hline \hline
\end{tabular}
\end{center}
\end{table}
\clearpage

\newpage
\begin{table}
\caption{The TO/LO optical phonon mode splitting of monoclinic LiN$_3$ and NaN$_3$ calculated within PBE+G06 functional.}
\label{tab.1}
\begin{center}
\begin{tabular}{ccc}\hline \hline
Compound& Mode & TO/LO \\ \hline
LiN$_3$& Au & 272.4/294.5  \\
      & Bu & 402.4/402.6 \\
      & Au & 604.4/612.2  \\
      & Bu & 617.6/625.3\\
      & Bu & 1994.4/2000.8 \\
NaN$_3$& Au & 203.1/213.1  \\
      & Bu & 260.2/265.5 \\
      & Au & 612.3/615.1  \\
      & Bu & 616.9/618.8\\
      & Bu & 1964.9/1974.3 \\ \hline
\end{tabular}
\end{center}
\end{table}
\end{document}